\documentclass[prd,onecolumn]{revtex4}
\pdfoutput=1
\usepackage{amssymb,latexsym}
\usepackage{amsmath,amsbsy,bbm}
\usepackage{ifpdf}
\usepackage{epsfig,bm}
\usepackage{graphicx,comment}
\usepackage{color}
\usepackage{soul}
\usepackage{mathtools}
\usepackage{comment}
\usepackage[normalem]{ulem}
\begin{document} 

\title{A Monte Carlo examination for the numerical values of universal quantities in spatial dimension two}
\author{Fu-Jiun Jiang*}
\affiliation{Department of Physics, National Taiwan Normal University,
88, Sec.4, Ting-Chou Rd., Taipei 116, Taiwan\\
\\
*Corresponding author. Email: fjjiang@ntnu.edu.tw}

\begin{abstract}
  By simulating a two-dimensional (2D) dimerized spin-1/2 antiferromagnet with 
	the quantum Monte Carlo method, the numerical values of two
  universal quantities associated with the quantum critical regime (QCR), namely
	$S(\pi,\pi)/\left(\chi_s T\right)$ and $c/\left(T\xi\right)$, are determined. Here $S(\pi,\pi)$, $\chi_s$, $c$, $\xi,$ and $T$ are the
  staggered structure factor, the staggered susceptibility, the spin-wave velocity, the correlation length, and the temperature, 
	respectively.
        For other QCR universal quantities, such as the Wilson ratio $W$ and $\chi_u c^2/T$ ($\chi_u$
        is the uniform susceptibility), it is shown that the addition of higher order theoretical contribution makes the agreement between the numerical and the
        analytic results worse.
  We find that the same scenario applies to $S(\pi,\pi)/\left(\chi_s T\right)$ and $c/\left(T\xi\right)$ as well.
  Specifically, our calculations lead to
  $S(\pi,\pi)/\left(\chi_s T\right)\sim 1.073$ and $c/\left(T\xi\right)\sim 0.963$ which are in better consistence with the leading
  theoretical predictions than those with the next-to-leading order terms.
  The presented outcome here as well as those in some relevant
   literature suggest that it is desirable to conduct a refinement of the analytic calculation to resolve the
   puzzle of why the inclusion of higher order terms leads to less accurate predictions for these universal quantities.
  
\end{abstract}

\maketitle

\section{Introduction}

For systems in spatial dimension two which have short-range interactions, 
the associated continuous symmetry cannot be broken
spontaneously (at finite temperature $T$)
due to the famous Mermin--Wagner
theorem \cite{Mer66}. Therefore, for these systems,
instead of a (finite-temperature) phase transition, an exotic
region called quantum critical regime (QCR) emerges.

Physical quantities related to QCR have been analytically calculated in detail
in Refs.~\cite{Chu93,Chu931,Chu94}. These quantities are universal, namely
their values only depend on the macroscopic, not the microscopic features
of the considered system(s). For two-dimensional (2D) quantum spin antiferromagnetic models,
these universal quantities include the Wilson ratio $W$, $\chi_u c^2/T$,
$S(\pi,\pi)/\left(\chi_s T\right)$, and $c/\left(\xi T\right)$. Here,
$\chi_u$, $c$, $S(\pi,\pi)$, $\chi_s$, and $\xi$ are the uniform susceptibility,
the spin-wave velocity, the staggered structure factor, the staggered susceptibility,
and the correlation length, respectively.

In Ref.~\cite{Chu94}, based on the effective $O(N)$ nonlinear sigma model, the numerical values for the
universal quantities mentioned in the previous paragraph have been determined to next-to-leading
order in the large $N$-expansion. The analytic calculations lead to $W \sim 0.1147$,
$\chi_u c^2/T \sim 0.27185$,
$S(\pi,\pi)/\left(\chi_s T\right) \sim 1.09$, and $c/\left(\xi T\right)\sim 1.04$.

Some of these quantities have been investigated using the first principle non-perturbative
quantum Monte Carlo simulations (QMC) in several studies \cite{Chu94,San95,Tro96,Tro97,Tro98,Kim00,Sen15,Tan181,Pen20,Jia23}.
Although earlier calculations
of Ref.~\cite{Chu94,San95,Tro98,Kim00} conclude that the Monte Carlo results agree well with the
theoretical predictions, later computations in Ref.~\cite{Sen15,Tan181,Pen20,Jia23} observe
non-negligible deviations between the QMC results and the analytic predictions.
In particular, it is found
in Ref.~\cite{Sen15,Tan181,Pen20,Jia23} that $W \sim 0.1248 (0.1238)$ and $\chi_u c^2/T \sim 0.32 (0.33)$.
Both the values of $W$ and $\chi_u c^2/T$, determined by the exact method, namely the QMC, differ 
from the associated analytic predictions by at least around 10 percent.

Interestingly, as pointed out in Ref.~\cite{Sen15}, with only the leading contributions,
the agreement between the analytic and the numerical outcomes for
$W$ and $\chi_u c^2/T$ is remarkably good.
Table 1 lists the theoretical leading results, up to the order $1/N$ outcomes as
well as the Monte Carlo determinations for $W$ and $\chi_u c^2/T$.
As can be seen from the table, the leading terms match the numerical results well, and
the addition of order $1/N$ corrections makes the agreement between the theoretical and the numerical values for $W$ and $\chi_u c^2/T$ worse.
It will be interesting to examine whether this scenario is systematic by computing other universal
quantities of QCR through an exact approach.

\begin{table}
	\label{tab1}
	\begin{center}
		\begin{tabular}{c|c|c|c}
			\hline
			{\text{Universal quantity}} & Leading order & Up to next-to-leading order & Monte Carlo \\
			\hline
			\hline
			$ W $ & 0.1243 & 0.1147 & 0.1248(3) (0.1238(3))\\
			\hline
			$ \chi_u c^2 /T $ & 0.3425 & 0.27185 & 0.33(1)\\
			\hline
			$ S(\pi,\pi)/\left(\chi_s T \right) $  & 1.076 & 1.09 & 1.0727(2) \\
                        \hline
                        $c/\left(T\xi\right)$ & 0.962424 & 1.0386 & 0.963(6) \\ 
                        \hline
			\hline
		\end{tabular}
	\end{center}
	\caption{The theoretical and Monte Carlo results for the numerical values of $W$, $\chi_u c^2/T$, $S(\pi,\pi)/\left(\chi_s T \right)$,
          and $c/\left(T\xi\right)$ (These values are from Refs.~\cite{Chu94,Sen15,Tan181,Pen20,Jia23} and present work).}
\end{table}

In this study, we perform large-scale Monte Carlo simulations to determine
the numerical values of $S(\pi,\pi)/\left(\chi_s T\right)$ and $c/\left(T\xi\right)$ of a 2D dimerized quantum
antiferromagnetic Heisenberg model. Dimerized quantum antiferromagnets and their spatially isotropic versions
have been
employed for many investigations regarding examining certain theoretical predictions,
such as the universality class of quantum phase transitions as well as the values of the low-energy constants from
the related effective field theories \cite{San951,Bea96,San97,Hog03,Wenzel08,Jiang09.1,Jin12,Jiang11.8,Yas13,Tan17}.

By extrapolating the data on linear system size $L=512$
with various temperatures, we arrive at $S(\pi,\pi)/\left(\chi_s T\right) \sim 1.073$.
Similar calculation leads to $c/\left(T\xi\right) \sim 0.963$.
Remarkably, like $W$ and $\chi_u c^2/T$, the theoretical leading
contributions for $S(\pi,\pi)/\left(\chi_s T\right)$ and $c/\left(T\xi\right)$ agree with our results
well, see table 1. The outcomes demonstrated in table 1 suggest strongly that
theoretical
leading terms have better agreement with the numerical values of the
universal quantities of QCR than those including the $1/N$ corrections is a
systematic trend.

It should be pointed out that the quantity $S(\pi,\pi)/\left(\chi_s T \right)$ associated with
the considered 2D dimerized quantum Heisenberg model has been studied in Ref.~\cite{Tan181}.
Here we simulate larger lattices and lower temperatures so that the bulk result can be
obtained with the systematic errors being (nearly) eliminated in principle. The $c/\left(T\xi\right)$ data shown here were obtained
sometime ago \cite{Tan181}, but are never published.

The presented outcomes in this study as well as those in Ref.~\cite{Sen15,Tan181,Pen20,Jia23}
suggest that it will be desirable to refine the associated analytic calculations
to resolve the puzzle of why the addition of hight order corrections leads to worse agreement between
the numerical and the theoretical
values for these universal quantities.

The rest of the paper is organized as follows. After the introduction,
the model and the measured observables are described in Sec.~II.
We then present the obtained results in Sec.~III.
Sec.~VI contains the conclusions of the present study.

\begin{figure}
  \vskip-0.5cm
       \includegraphics[width=0.325\textwidth]{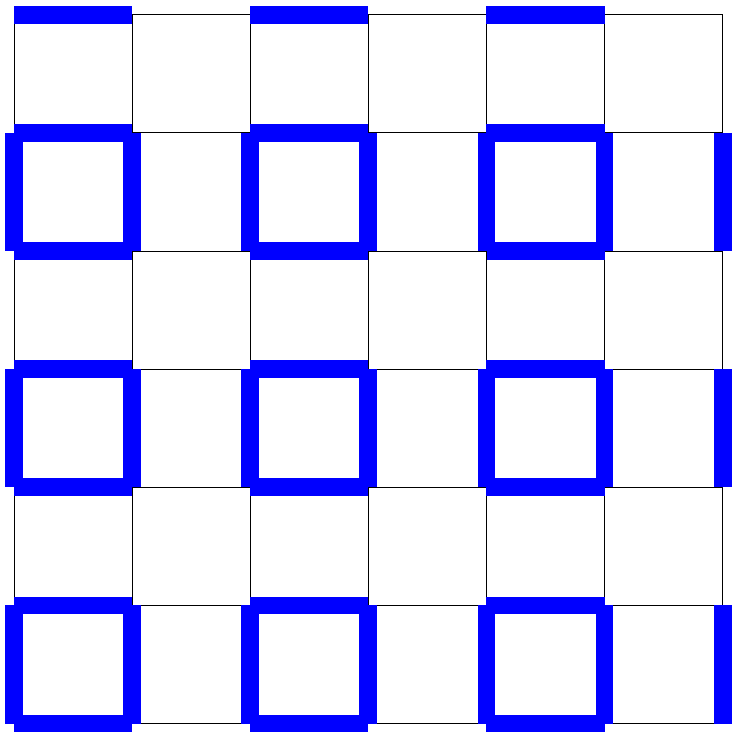}        
        \vskip-0.2cm
        \caption{The plaquette model considered in this study. The thick
          and thin bonds represent the couplings of strength $J_2$ and $J_1$,
          respectively. The figure is from Ref.~\cite{Tan181}.}
        \label{fig0}
\end{figure}
  
\section{The considered model and observables}

The Hamiltonian of the studied 2D quantum
dimerized plaquette Heisenberg model has the form
\begin{eqnarray}
\label{hamilton}
H &=& \sum_{\langle ij \rangle}J_1\,\vec S_i \cdot \vec S_{j} 
+ \sum_{\langle i'j' \rangle}J_2\,\vec S_{i'} \cdot \vec S_{j'}, 
\end{eqnarray}
where $J_1$ and $J_2$ are the antiferromagnetic
couplings connecting nearest neighbor spins $\langle  ij \rangle$
and $\langle  i'j' \rangle$, respectively,
and $\vec{S}_i$ is the spin-$\frac{1}{2}$ operator at site $i$.
Fig.~\ref{fig0} shows the graphical representation of the considered model.
In this study, we let $J_1$ be 1 and vary the value of $J_2$ so that
a phase transition from the ordered phase to the disordered order takes place
at a particular value of $J_2 > J_1$. It is known in the literature that the
value of $J_2/J_1$ where the mentioned phase transition occurs is $J_2/J_1 = 1.8230(2)$ \cite{Wen09} (This particular value is typically denoted by $(J_2/J_1)_c$ in the literature).

To obtain a high precision value for the universal quantity $S(\pi,\pi)/\left(\chi_s T\right)$ and $c/\left(T\xi\right)$ of QCR,
the staggered structure factor $S(\pi,\pi)$, the
staggered susceptibility $\chi_s$, the correlation length $\xi$, and spin-wave velocity $c$ are calculations from the QMC. 
On a (squared) system with linear size $L$ and inverse temperature $\beta$, $S(\pi,\pi)$ and $\chi_s$ are defined as
\begin{eqnarray}
  S(\pi,\pi) &=& 3 L^2 \left\langle ( m_s^z )^2\right\rangle\\
  \chi_s &=& 3L^2\int_0^{\beta} \langle m_s^z(\tau) m_s^z(0)\rangle d\tau,
\end{eqnarray}
where $ m_s^z = \frac{1}{L^2}\sum_{i}^{N}(-1)^{i_i + i_2}S_i^z$ with $N = L^2$. In addition, $\xi$ is given by $\xi = 0.5(\xi_1 + \xi_2)$,
where $\xi_1$ and $\xi_2$ are defined as
\begin{eqnarray}
\xi_1 &=& \frac{L}{2\pi}\sqrt{\frac{S(\pi,\pi)}{S(\pi+2\pi/L,\pi)}-1} \nonumber \\
\xi_2 &=&   \frac{L}{2\pi}\sqrt{\frac{S(\pi,\pi)}{S(\pi,\pi+2\pi/L)}-1},
\end{eqnarray}
here the quantities $S(\pi+2\pi/L,\pi)$ and $S(\pi,\pi+2\pi/L)$ are the
Fourier modes with the second largest magnitude.
Finally, the spin-wave velocity $c$ is obtained
through the temporal and spatial winding numbers squared
($\langle W_t^2\rangle$ and $\langle W_i^2 \rangle$ with $i\in\{1,2\}$).

\section{Numerical Results}

To calculate the numerical values of $S(\pi,\pi)/\left(\chi_s T\right)$ and $c/\left(T\xi\right)$ in an exact approach,
we have performed large-scale
quantum Monte Carlo calculations (QMC) using the stochastic series
expansion algorithm (SSE) with efficient operate-loop update \cite{San99}.
The outcomes obtained are detailed below.

\subsubsection{$S(\pi,\pi)/\left(\chi_s T\right)$}

\begin{figure}
  \vskip-0.5cm
       \includegraphics[width=0.5\textwidth]{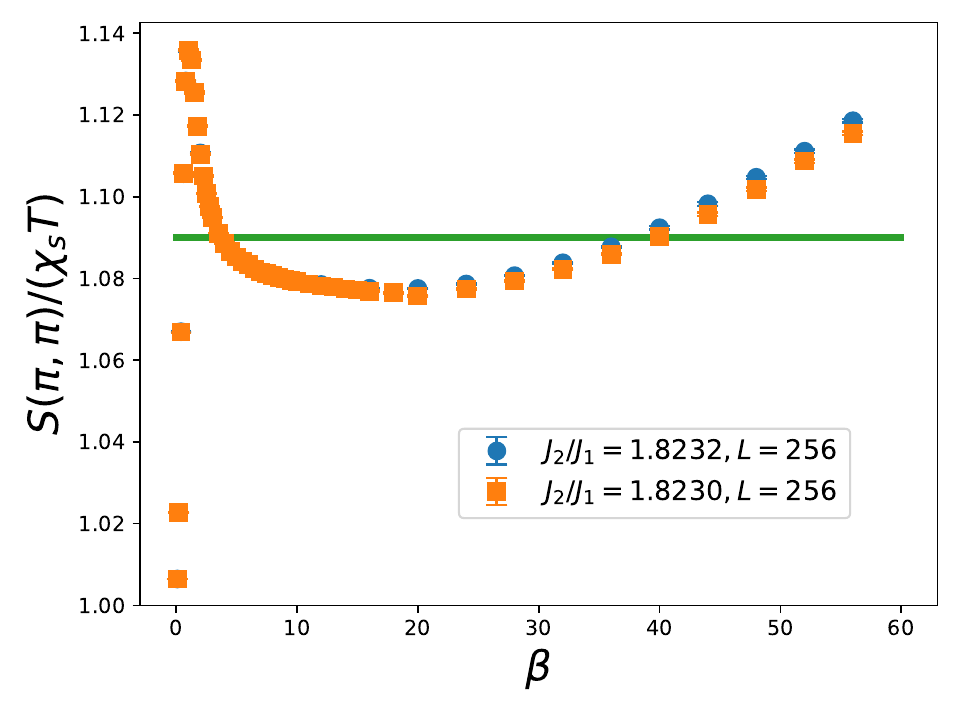}        
        \vskip-0.2cm
        \caption{$S(\pi,\pi)/\left(\chi_s T\right)$ as functions of $\beta$ for $J_2/J_1$ = 1.8230 and 1.8232. Part of the data are from Ref.~\cite{Tan181}.}
        \label{fig1}
\end{figure}

It is known that the critical point $(J_2/J_1)_c$ for the chosen dimerized quantum antiferromagnetic model is given by 1.8230(2) \cite{Wen09}.
Therefore, for $L=256$ we have carried out calculations with $J_2/J_1$ = 1.8230 and 1.8232.
The resulting $S(\pi,\pi)/\left(\chi_s T\right)$ as functions of $\beta$
are shown in fig.~\ref{fig1}. As can be seen in the figure, the deviation of the values for $S(\pi,\pi)/\left(\chi_s T\right)$
due to the variant of $(J_2/J_1)_c$ within the error bar is very mild. As a result, we have conducted the investigation at $J_2/J_1$ = 1.8230.

\begin{figure}
  \vskip-0.5cm
       \includegraphics[width=0.5\textwidth]{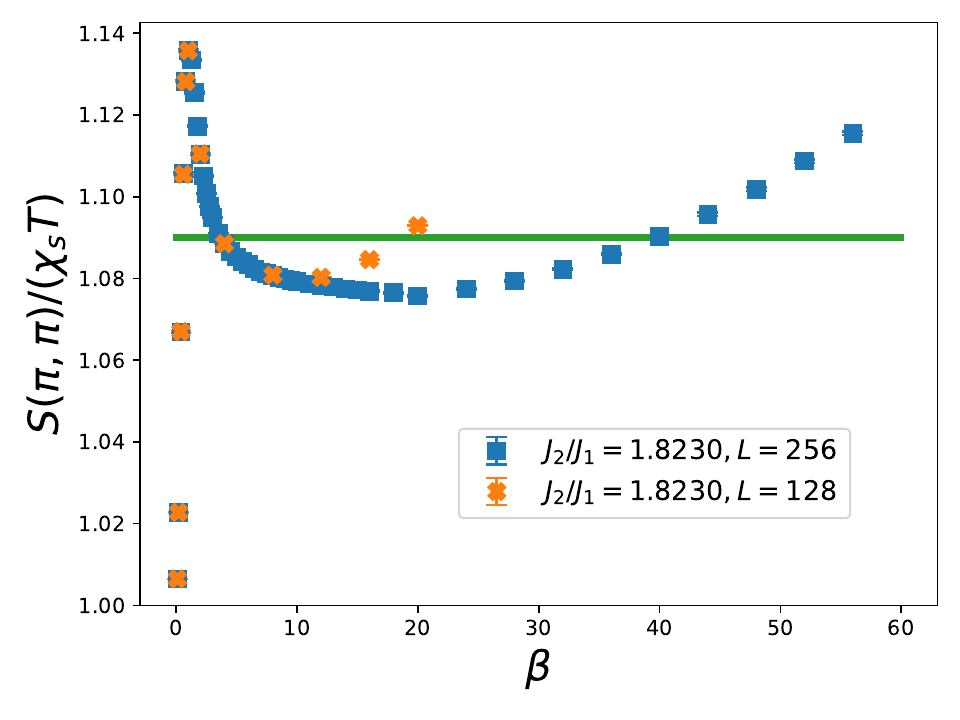}        
        \vskip-0.2cm
        \caption{$S(\pi,\pi)/\left(\chi_s T\right)$ as functions of $\beta$ at $J_2/J_1$ = 1.8230 for $L=128$ and $L=256$. Part of the data are from Ref.~\cite{Tan181}.}
        \label{fig2}
\end{figure}

\begin{figure}
  \vskip-0.5cm
       \includegraphics[width=0.5\textwidth]{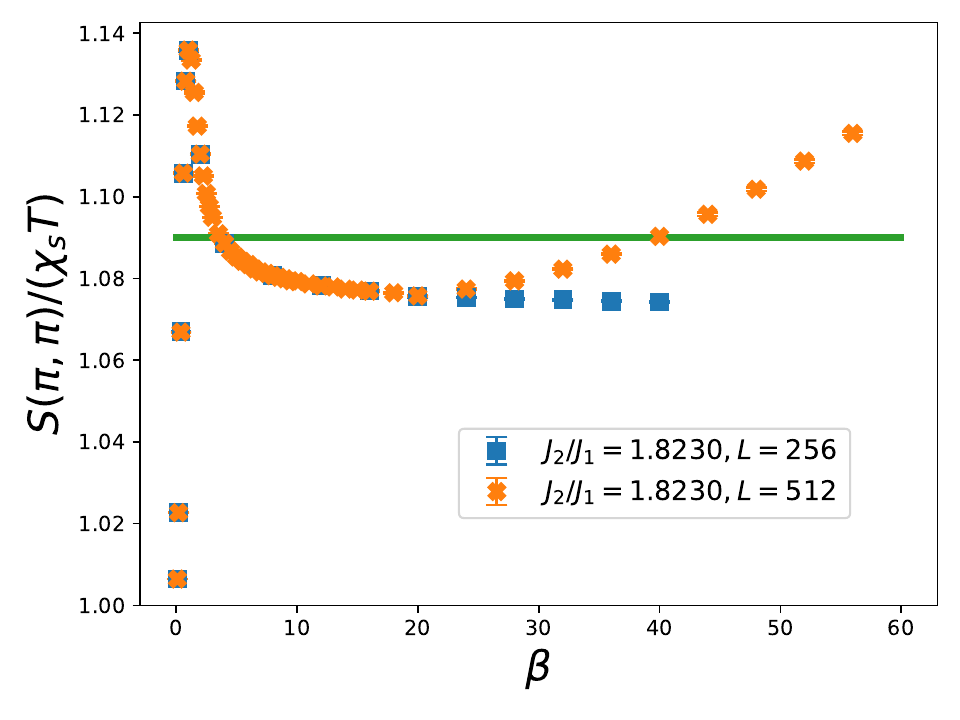}        
        \vskip-0.2cm
        \caption{$S(\pi,\pi)/\left(\chi_s T\right)$ as functions of $\beta$ at $J_2/J_1$ = 1.8230 for $L=256$ and $L=512$. Part of the data are from Ref.~\cite{Tan181}.}
        \label{fig3}
\end{figure}

Figure \ref{fig2} shows $S(\pi,\pi)/\left(\chi_s T\right)$ as functions of $\beta$ for $L=128$ and $L=256$. The results in the figure
imply that the finite-size effect starts to become serious when the data curve associated with the smaller $L$ begins to rise with $\beta$.
A similar trend is also observed for the data associated with $L=256$ and $L=512$, see fig.~\ref{fig3}.

\begin{figure}
  \vskip-0.5cm
       \includegraphics[width=0.5\textwidth]{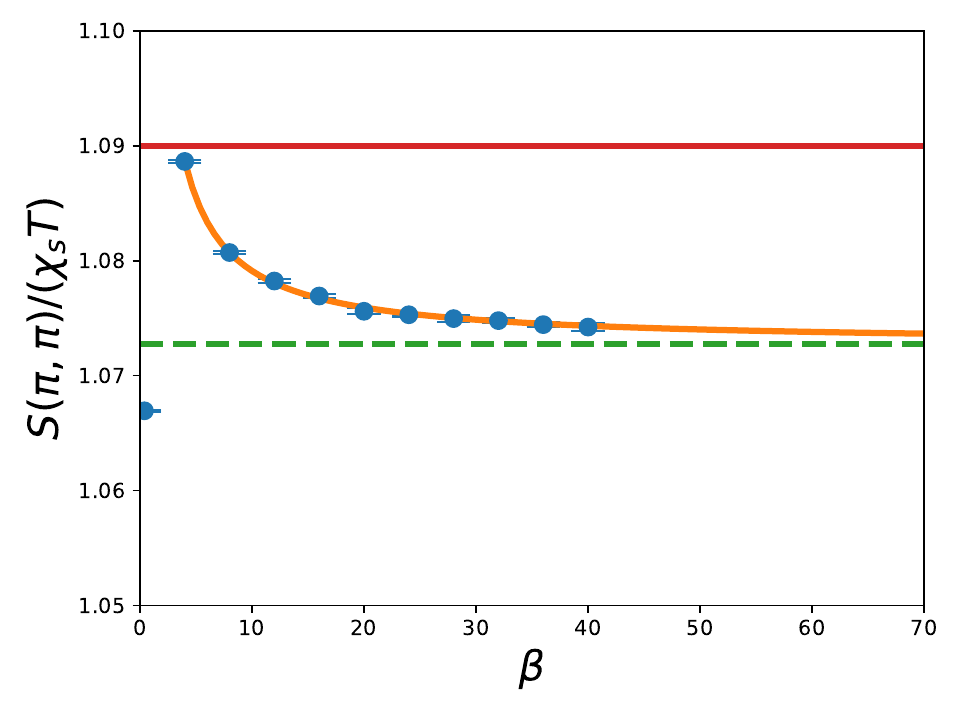}        
        \vskip-0.2cm
        \caption{$S(\pi,\pi)/\left(\chi_s T\right)$ as a function of $\beta$ at $J_2/J_1$ = 1.8230 for $L=512$.
          The solid curve is obtained by using the results of the fits. The horizontal solid and dashed lines represent
        1.09 (up to next to leading order prediction) and 1.0727 (leading order prediction), respectively.}
        \label{fig4}
\end{figure}

Based on the observation described above and shown in figs.~\ref{fig2} and \ref{fig3}, it is beyond doubt that the deviation of
$L = 512$ data from the corresponding bulk ones is insignificant since the trend of $L=512$ data still moves downward for the used values of $\beta$. By applying the ansatz $a + b/\beta$ to fit the $L=512$ data with $\beta > 4.0$ leads to $a =1.0727(2)$. With this fitting
result one arrives at $S(\pi,\pi)/\left(\chi_s T\right) \sim 1.0727(2) \sim 1.073$. Figure \ref{fig4} demonstrates the $L=512$ data
of $S(\pi,\pi)/\left(\chi_s T\right)$ as a function of $\beta$. The solid curve appearing in the figure is obtained by using the fitting results.
Finally, the horizontal solid and dashed lines are 1.09 and 1.0727, respectively.

\subsubsection{$c/\{\left(T\xi\right)$}

The calculation of $c/\left(T\xi\right)$ requires the numerical value of $c$. The quantity $c$ can be determined by
the temporal and spatial winding numbers squared $\langle W_t^2\rangle$ and
$\langle W^2 \rangle = 0.5\left(\langle W_1^2 \rangle + \langle W_2^2 \rangle\right)$ \cite{Jia11}. Specifically,
for a given $L$, one varies $\beta$ so that $\langle W_t^2\rangle$ and $\langle W^2 \rangle$ take about the same values.
When this condition is met, the $c(L)$ associated with the given $L$ is $c(L) = L/\beta$. With the obtained
$c(L)$ on several finite lattices, one can determine the bulk $c$ by extrapolating the data of $c(L)$.

\begin{figure}
  \vskip-0.5cm
       \includegraphics[width=0.5\textwidth]{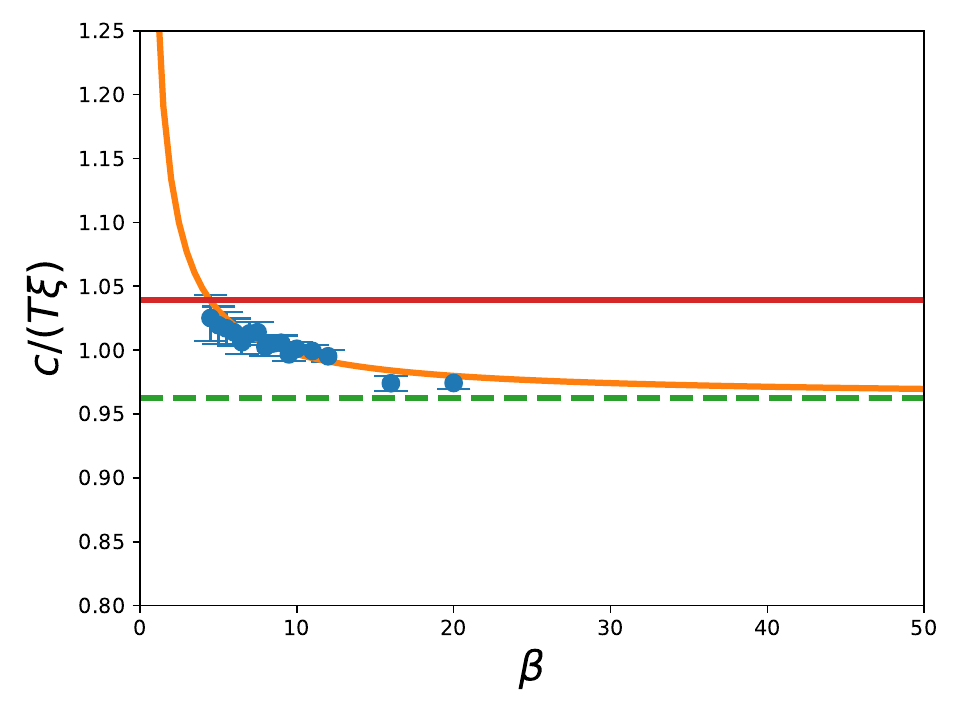}        
        \vskip-0.2cm
        \caption{$c/\left(T\xi\right)$ as a function of $\beta$ at $J_2/J_1$ = 1.8230.
          The solid curve is obtained by using the results of the fits. The horizontal solid and dashed lines represent
        1.04 (up to next to leading order prediction) and 0.962424 (leading order prediction), respectively.}
        \label{fig5}
\end{figure}

The spin-wave velocity required for determining $c/\{\left(T\xi\right)$ has been calculated in Ref.~\cite{Tan181} and
has a value of 2.163(4). $c/\left(T\xi\right)$ as a function of $\beta$ is depicted in fig.~\ref{fig5}. For each value
of $\beta$, the associated $c/\left(T\xi\right)$ is either the bulk one or has nearly negligible finite-size effect.
Therefore, conducting fits using the data shown in the figure should lead to reliable zero temperature result for
$c/\left(T\xi\right)$. With fits by employing the ansatz $a + b/\beta$, we arrive at $a = 0.963(6)$ which is
the zero temperature value for $c/\left(T\xi\right)$. The obtained $c/\left(T\xi\right) = 0.963(6)$ agrees reasonably
well with the leading order analytic prediction $0.962424$, but deviates from the result of adding next-to-leading
order contribution $1.04$.

\begin{figure}
  \vskip-0.5cm
       \includegraphics[width=0.5\textwidth]{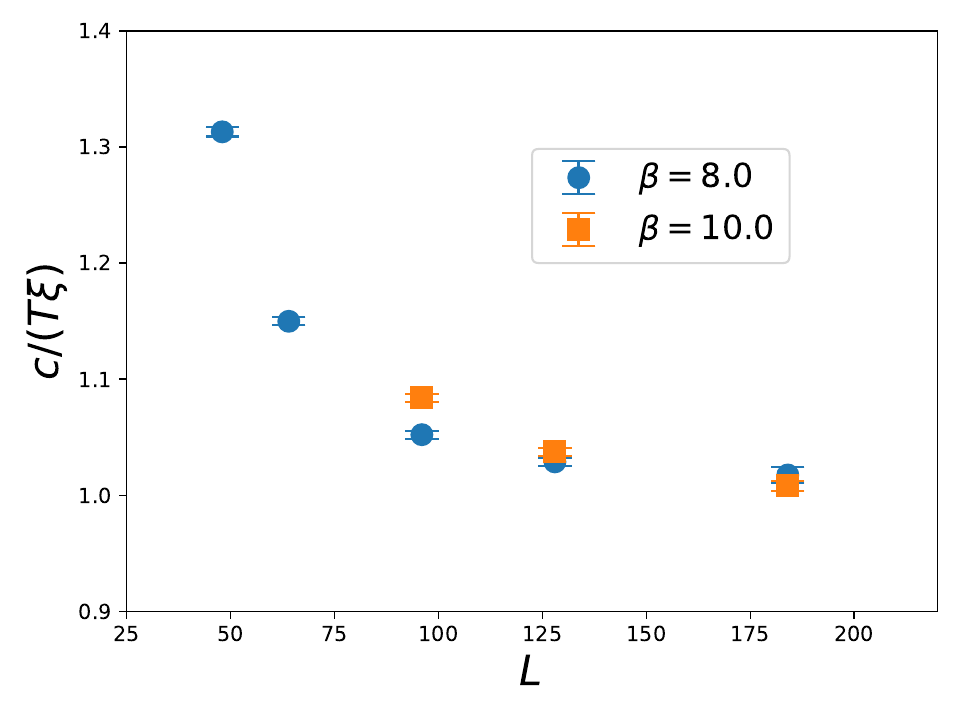}        
        \vskip-0.2cm
        \caption{$c/\left(T\xi\right)$ as a function of $L$ at $J_2/J_1$ = 1.8230 for
          two values of inverse temperature $\beta = 8.0$ and $\beta = 10.0$.}
        \label{fig6}
\end{figure}

Fig.~\ref{fig6} shows the $L$-dependence of $c/\left(T\xi\right)$ for two values of inverse temperature $\beta = 8.0$ and $\beta = 10.0$.
As can be seen from the figure, the magnitude of $c/\left(T\xi\right)$ decreases as $\beta$ increases. Hence, if the
the data shown in fig.~\ref{fig5} are slightly away from their bulk ones, then one expects that the bulk data curve will lower a little bit. 
This scenario will lead to the same conclusion as claimed in the previous paragraph. Specifically, the leading analytic prediction
matches better with the numerical result than the one including the next-to-leading order contribution.

\subsubsection{$\chi_u c^2/T$ and $W$}

For completeness, we reproduce the results associated with $\chi_u c^2/T$ and $W$ using the data of Refs.~\cite{Jia23} and \cite{Pen20}. These
data are obtained by simulating the 2D dimerized Herringbone quantum Heisenberg model.

\begin{figure}
	\vskip-0.5cm
	\vbox{
	\includegraphics[width=0.45\textwidth]{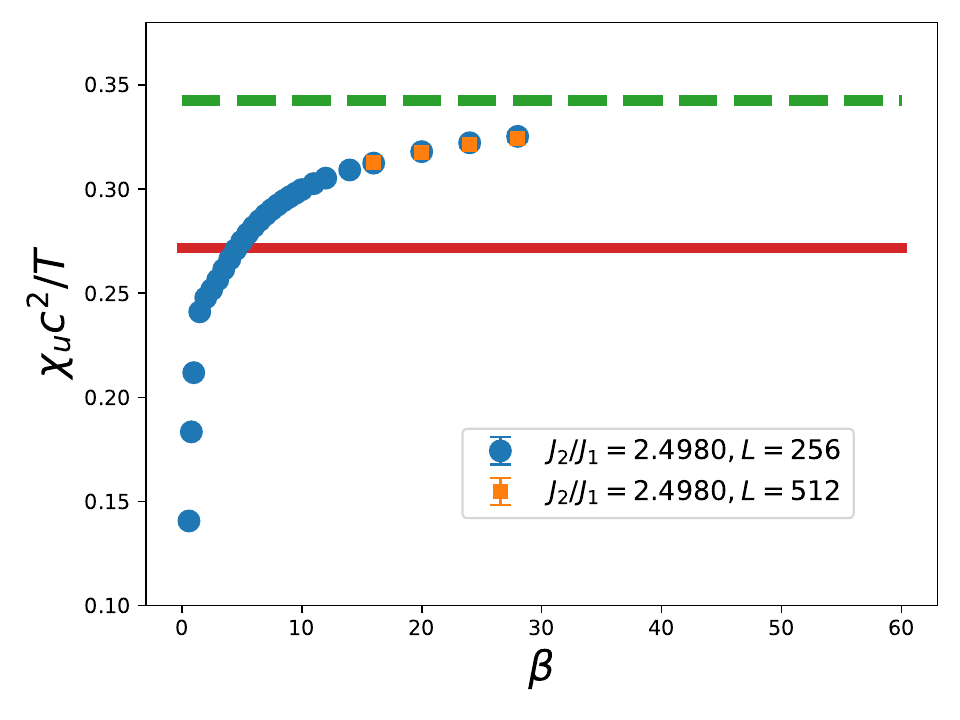}
	 \includegraphics[width=0.45\textwidth]{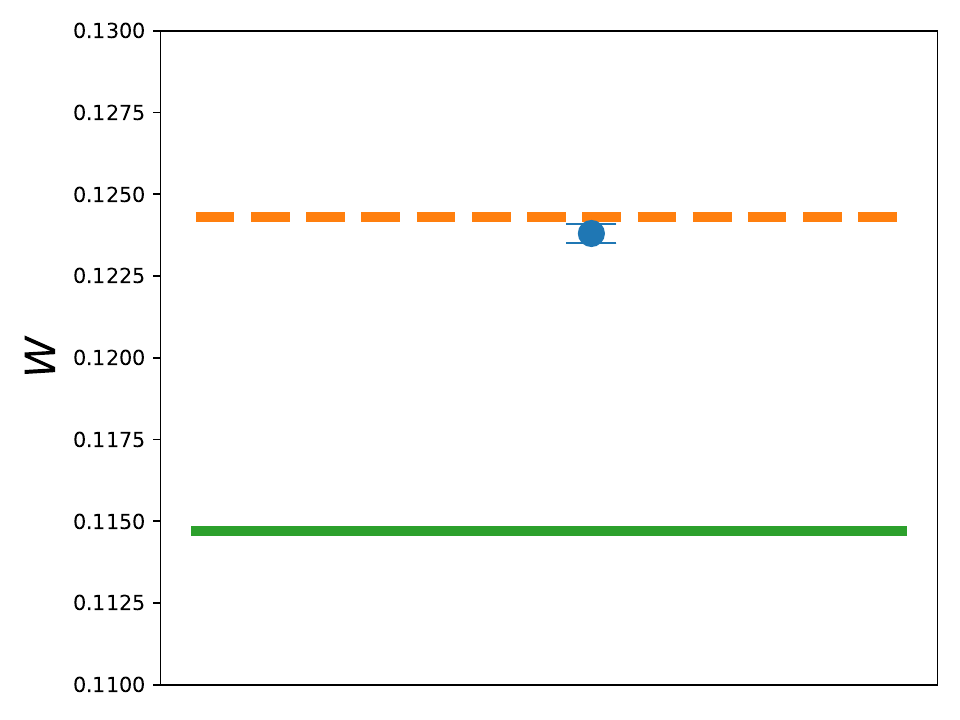} 
}      
	\vskip-0.2cm
	\caption{(Left) $\chi_u c^2/T$ as 
		functions of the inverse temperature $\beta$ \cite{Jia23}. (Right) The value of Wilson ratio $W$ ($W = 0.1238(3))$ \cite{Pen20}. In both panels, the dashed and solid lines represent the leading order and up to next-to-leading order theoretical predictions, respectively.}
	\label{fig7}
\end{figure}

The left panel of fig.~\ref{fig7} is the quantity $\chi_u c^2/T$ as 
functions of the inverse temperature $\beta$ calculated in Ref.~\cite{Jia23}. The dashed and solid lines in the figure are the leading order and up to next-to-leading order theoretical predictions. It is clear from 
the figure that
the leading order result has better match with the numerical data than that including the next-to-leading order contribution.

Finally, the right panel of fig.~\ref{fig7} is the Wilson ratio $W$ obtained in Ref.~\cite{Pen20} ($W=0.1238(3)$). Similar to the left panel of fig.~\ref{fig7}, the dashed and solid lines in the figure represent the leading order and up to next-to-leading order analytic outcomes. It is beyond doubt from 
the figure that the agreement between
the numerical data and the leading order result is much better than that when one compares the data with the prediction including the next-to-leading order contribution.

\section{Discussions and Conclusions}

In this study, we simulate the 2D dimerized quantum plaquette Heisenberg model using
QMC. In addition, fitting the relevant data obtained on large lattices and at moderate low temperatures,
we find that the numerical values of $S(\pi,\pi)/\left(\chi_s T\right)$ and $c/\left(T\xi\right)$ are given by
$S(\pi,\pi)/\left(\chi_s T\right) \sim 1.073$ and $c/\left(T\xi\right) \sim 0.963$.

Based on the presented outcomes here and those in Refs.~\cite{Sen15,Pen20,Jia23},
it is unexpected that for these four quantities $W$, $\chi_u c^2/T$, $S(\pi,\pi)/\left(\chi_s T\right)$, and $c/\left(T\xi\right)$ of QCR,
the leading large-$N$ expansion
calculations lead to better agreement between the theoretical and the numerical values than the ones including
the $1/N$ corrections. It will be extremely interesting to understand whether this trend can be explained rigorously from a analytic point of view. 
Finally, since in all these cases, the $N$ in the large $N$-expansion calculations is 3 which is not large, it is also likely that a calculation
to the next-to-next-to-leading order may coincide the
results of Monte Carlo with those of analytic computations.

In conclusion, it is desirable to conduct further theoretical
calculations to resolve the puzzle of the numerical and the theoretical results for these universal quantities shown in this study and
in Refs~\cite{Sen15,Pen20,Jia23}.


\section*{Acknowledgement}
We thank D.-R.~Tan and J.-H. Peng for their involvement in Refs.~\cite{Tan181} and \cite{Pen20}.

\section*{Funding}
Partial support from National Science and Technology Council (NSTC) of
Taiwan is acknowledged (Grant numbers: NSTC 112-2112-M-003-016- and NSTC 113-2112-M-003-014-). 


\section*{Conflict of Interest}
The author declares no conflict of interest.

\section*{Data Availability Statement}
Data are available from the corresponding author
on reasonable request.


\begin{thebibliography}{1}

\bibitem{Mer66}
  N. D. Mermin and H. Wagner, 
    Phys. Rev. Lett. {\bf 17}, (1966) 1133. 

\bibitem{Chu93}
A.~V.~Chubukov and S.~Sachdev, Phys. Rev. Lett. {\bf 71}, 169 (1993).

\bibitem{Chu931}
A. V. Chubukov and S. Sachdev, Phys. Rev. Lett. {\bf 71}, 2680 (1993).
    
\bibitem{Chu94}
  A.~V.~Chubukov, S.~Sachdev, and J.~Ye, 
    Phys. Rev. B {\bf 49}, (1994) 11919. 

\bibitem{San95}
A.~W.~Sandvik, A.~V.~Chubukov, and S.~Sachdev, Phys. Rev. B {\bf 51}, 16483 (1995)

\bibitem{Tro96}
M.~Troyer, H.~Kantani, and K.~Ueda, Phys.~Rev.~Lett. {\bf 76}, 3822 (1996).

\bibitem{Tro97}
Matthias Troyer, Masatoshi~Imada, and Kazuo~Ueda, J. Phys. Soc. Jpn. 66, 2957 (1997).

    
\bibitem{Tro98}
  Jae-Kwon Kim and Matthias~Troyer, 
  Phys.~Rev.~Lett. {\bf 80}, (1998) 2705. 
  
\bibitem{Kim00}
Y.~J.~Kim and R.~J.~Birgeneau, Phys. Rev. B {\bf 62}, 6378 (2000).
  
\bibitem{Sen15}
  A. Sen, H. Suwa, and A. W. Sandvik, 
  Phys. Rev. B {\bf 92}, (2015) 195145. 

\bibitem{Tan181}
  D.-R. Tan and F.-J. Jiang, 
  Phys. Rev. B {\bf 98}, (2018) 245111. 

\bibitem{Pen20}
J.-H. Peng, D.-R. Tan, and F.-J. Jiang, Phys. Rev. B {\bf 102}, 214206 (2020). 

\bibitem{Jia23}
Fu-Jiun Jiang, Results in Physics 44 (2023) 106119.  

\bibitem{San951}
  A.~W.~Sandvik and M.~Vekic, 
  Phys. Rev. Lett. {\bf 74}, 1226 (1995). 
  
\bibitem{Bea96}
B.~B.~Beard and U.-J. Wiese,
Phys. Rev. Lett. {\bf 77}, 5130 (1996). 

\bibitem{San97}
  A.~W.~Sandvik, Phys. Rev. B {\bf 56}, 11678 (1997). 

\bibitem{Hog03} 
K. H. H\"oglund and A. W. Sandvik, 
Phys. Rev. Lett. {\bf 91},
077204 (2003). 

  
\bibitem{Wenzel08}
S.~Wenzel, L.~Bogacz, and W.~Janke, 
  Phys. Rev. Lett. {\bf 101}, 127202 (2008). 
  
\bibitem{Jiang09.1}
F.-J. Jiang, F. K\"ampfer, and M. Nyfeler, 
Phys. Rev. B {\bf 80}, 033104 (2009). 
\bibitem{Jin12}
  S.~Jin and A.~W.~Sandvik, 
  Phys. Rev. B {\bf 85}, 020409(R) (2012). 

  \bibitem{Jiang11.8}
    F.-J. Jiang, 
    Phys. Rev. B {\bf 85} 014414 (2012). 

\bibitem{Yas13}
Shinya Yasuda and Synge Todo, 
Phys. Rev. E {\bf 88} 061301 (2013). 
  
\bibitem{Tan17}
D.-R. Tan and F.-J. Jiang, 
Phys. Rev. B {\bf 95}, 054435 (2017). 


\bibitem{Wen09}
S. Wenzel and W. Janke, Phys. Rev. B {\bf 79}, 014410 (2009).


\bibitem{San99}
  A.~W.~Sandvik, 
  Phys. Rev. B {\bf 55}, (1999) R14157. 

\bibitem{Jia11}
F.-J. Jiang,
Phys. Rev. B {\bf 83}, (2011) 024419. 


\end{thebibliography}
\end{document}